\documentclass[10pt,twocolumn,twoside]{IEEEtranTCOM}
\ifCLASSINFOpdf
\else
\fi
\usepackage{graphicx}
\usepackage{enumitem}
\usepackage{amsmath}
\usepackage{algpseudocode}
\usepackage{amssymb}
\usepackage{times}
\usepackage{endnotes}
\usepackage{stfloats}
\usepackage{stkernel}
\usepackage{color}
\usepackage{enumerate}
\usepackage{array}
\usepackage{relsize}
\usepackage[belowskip=-10pt,aboveskip=3pt,footnotesize]{caption}
\usepackage{cite}
\renewcommand{\baselinestretch}{0.98}

\newcommand{\HRule}{\rule{\linewidth}{0.3mm}}

\newcommand{\refeq}[1]{(\ref{#1})}
\newcommand{\tr}[1]{\mathrm{#1}}

\begin{document}

\title{Constellation Optimization in the Presence of Strong Phase Noise}

\author{Rajet~Krishnan,~\IEEEmembership{Student~Member,~IEEE,}
        Alexandre~Graell~i~Amat,~\IEEEmembership{Senior~Member,~IEEE,}
        Thomas~Eriksson,
        and~Giulio~Colavolpe, ~\IEEEmembership{Senior~Member,~IEEE}

\thanks{Rajet Krishnan, Alexandre Graell i Amat and Thomas Eriksson are with the Department
of Signals and Systems, Chalmers Univeristy of Technology, Gothenburg, Sweden (e-mail: \{rajet,
alexandre.graell, thomase\}@chalmers.se).}%
\thanks{Giulio Colavolpe is with Dipartimento di Ingegneria dell'Informazione,  University of Parma, Parma, Italy  (e-mail: giulio@unipr.it).}
\thanks{Research supported by the Swedish Research Council under grant \#2011-5961.}

}%

\markboth{IEEE Transactions on Communications}%
{Accepted for Publication}

\maketitle% \thispagestyle{empty}

\begin{abstract}
In this paper, we address the problem of optimizing signal constellations for strong phase noise. The problem is investigated by considering three optimization formulations, which provide an analytical framework for constellation design. In the first formulation, we seek to design constellations that minimize the symbol error probability (SEP) for an approximate ML detector in the presence of phase noise. In the second formulation, we optimize constellations in terms of mutual information (MI) for the effective discrete channel consisting of phase noise, additive white Gaussian noise, and the approximate ML detector. To this end, we derive the MI of this discrete channel. Finally, we optimize constellations in terms of the MI for the phase noise channel. We give two analytical characterizations of the MI of this channel, which are shown to be accurate for a wide range of signal-to-noise ratios and phase noise variances.  For each formulation, we present a detailed analysis of the optimal constellations and their performance in the presence of strong phase noise. We show that the optimal constellations significantly outperform conventional constellations and those proposed in the literature in terms of SEP, error floors, and MI.

%\textit{quadrature-amplitude modulation} (QAM), \textit{phase-shift keying} (PSK) ;In all the three formulations, a strict average power constraint is used, and equally likely symbols are assumed.

\end{abstract}

\section{Introduction}
\label{sec:intro}

Recent times have witnessed a tremendous surge in data rate requirements in wireless networks \cite{boch09}. In this context, wireless ethernet has become a preferred choice for backhaul connectivity. In order to achieve high data rates over wireless links, high order signal constellations are being considered for transmission, where phase noise impairments can incur heavy losses in terms of error rate and throughput \cite{boch09}. These impairments have to be appropriately compensated to enhance system performance.

The problem of compensating systems affected by phase noise to achieve near-coherent performance has been studied extensively in the past, e.g., \cite{marc99,meng97} and references therein. One of the most widely used techniques is to design joint detection and estimation algorithms at the receiver \cite{raj12,raj11,fos73,kam94,cola05,noel07,vb09}. In these approaches, standard error correcting codes \cite{lin} are used along with conventional phase shift keying (PSK) or quadrature amplitude modulation (QAM) constellations for transmitting data over the channel. Designing  error correcting codes that are suitable for phase noise scenarios is also an effective method for combatting these impairments \cite{ferr12,karu08,cola08,colavolpe11,imai04}.

To further improve the performance, a properly designed constellation can also be employed. The problem of arranging $M$ points on a two-dimensional plane such that a target objective function is optimized is a classical problem in communication theory \cite{kern72}. For decades, this problem has been studied for different channel models \cite{fos74,lot10,pfau10,chris12,chris13}. Specifically in the context of phase noise over a wireless link, the design of constellations that optimize the symbol error probability (SEP) was first rigorously addressed by Foschini \textit{et al.} in \cite{fos73}. In their work, an approximate ML detector was derived for a memoryless phase noise channel, and constellations that optimize its (approximate) SEP were obtained. In \cite{sam98}, constellations robust to phase noise were constructed heuristically such that they have a low decoding complexity. The shaping gain that can be achieved over conventional PSK and QAM constellations by properly designing constellations for phase noise was studied in \cite{belzer02}. The work in \cite{hani12} investigated the performance of high order amplitude-phase shift keying (APSK) constellations as compared to the conventional PSK and QAM constellations. In  \cite{li08}, the SEP of the ML detector for a given phase offset was derived, and this criterion was minimized for designing constellations. In \cite{kwa08}, a simple method for constructing spiral QAM constellations was presented, and their performances were compared with those of other conventional constellations in the presence of phase noise. In a more recent paper \cite{tor12}, the problem of designing constellations that maximize the mutual information (MI) of a memoryless phase noise channel was addressed. Here, the (approximate) MI for the channel was derived, and optimal constellations were obtained by maximizing the MI using a simulated annealing algorithm.

Prior work has demonstrated that constellations designed for phase noise scenarios help gain substantially in terms of SEP and MI compared to conventional constellations. However, in most prior work (except \cite{fos73} and \cite{tor12}) ad-hoc methods have been used. There has been very limited effort to address this problem based on rigorous optimization formulations in terms of SEP or MI. This can be partly attributed to the challenge in analytically deriving the exact ML detector for this problem, its SEP \cite{raj12}, \cite{kam09} and the MI for a phase noise channel \cite{tor12}. %These factors motivate the need to revisit the problem of constellation design based on optimization formulations that use accurate/approximate target objective functions like SEP or MI.

%Constellations that minimize SEP are desirable in uncoded systems. Optimizing constellations such that they maximize MI enhance performances of coded systems, especially for low signal-to-noise ratios (SNRs) \cite{fitz07}.

{In this work, we present three optimization formulations that provide an analytical framework for designing constellations in the presence of phase noise. In the first formulation, we design constellations to minimize the SEP of an approximate ML detector for a memoryless phase noise channel, derived in \cite{raj12}. Constellations that minimize SEP for the phase noise channel are desirable in uncoded systems. Also, there are latency limited systems and applications such as coordination of base stations in 4G cellular networks and feedback loops in control systems that are preferably uncoded.  In coded systems, some levels of processing such as clock recovery, forward error correcting (FEC) frame preamble decoding, and adaptive equalization are based on the SEP performance.}

{In the next formulation, we design constellations that maximize the MI of the effective (discrete) channel consisting of the memoryless phase noise, additive white Gaussian noise (AWGN) and the ML detector. The MI of this discrete channel is derived based on the ML detector in \cite{raj12}. This MI is an interesting performance metric for systems that employ codes with hard decision decoding, like Reed Solomon codes \cite{lin}. Hard decision decoding, though generally resulting in lower performance than soft decoding, incurs much lower decoder complexity \cite{swaszek98}.}

{For the final formulation, we seek to design constellations that maximize the MI of a memoryless phase noise channel. To this end, we present two new analytical characterizations of the MI for this channel based on (i) a likelihood function as derived in \cite{raj12}, using a high SNR approximation, and (ii) a likelihood function, newly derived in this paper, based on a low instantaneous phase noise approximation. Unlike the MI derived in \cite{tor12}, these characterizations are analytically simpler and accurate for a wide range of SNRs and phase noise variances. The MI for this channel is relevant for soft-decoding. It gives an upper bound on the achievable rate for any decoder \cite{fitz07}, and is particularly relevant for symbol-based decoders such as in trellis-coded modulation or LDPC-based nonbinary coded schemes, and for systems that employ binary capacity-achieving codes like multilevel codes \cite{mlcm98}.
By properly designing non-binary codes to match the optimized constellations, or using binary multilevel codes, the MI of the constellation can be approached.
%Furthermore, MI-optimized constellations can be directly used with appropriately designed non-binary codes (i.e., codes designed with respect to the constellation), and the maximum achievable rate of the system corresponds to the MI of the constellation. This also applies to binary capacity-achieving like multilevel codes.
}

{In coded systems that employ bit-interleaved coded modulation (BICM), the constellation and its labeling must be jointly optimized in order to maximize the generalized mutual information (GMI) \cite{tor12}. The GMI is strictly upper-bounded by the MI of the channel and the bound is tight for asymptotically-low and high SNR scenarios. Addressing the case of BICM systems for other SNRs would require the derivation of the GMI for the phase noise channel under consideration, which is beyond the scope of this paper. However we remark that it is possible to optimize the labeling  for a given optimized constellation, and this two-stage approach, though sub-optimal, can increase the GMI of the system.}

For the three formulations, we find the optimized constellations, and analyze their performances in the presence of strong phase noise. We show that the optimized constellations outperform conventional constellations and those proposed in the literature in terms of SEP, error floors and MI. As expected, the optimized constellations do not possess any particular structure, which may make their practical implementation cumbersome. In order to circumvent this difficulty, we also design APSK constellations as in \cite{chris12,chris13}, and optimize them in terms of their SEP and MI performances.

The remainder of the paper is organized as follows. In Section \ref{sec:system_model}, we discuss the system model. In Section \ref{sec:analysis}, we derive a new likelihood function and detector using a low instantaneous phase noise approximation for a memoryless phase noise channel. In Section \ref{sec:metrics}, we discuss several performance metrics for the channel under consideration. In Section \ref{sec:cons_opt}, we present the optimization formulations that are used to design constellations. The optimization formulation to design structured APSK constellations is presented in Section \ref{sec:ring_const}. In Section \ref{sec:results} we compare the SEP and MI of different constellations, and we conclude our work and highlight the key findings in Section \ref{sec:conc}.

Notation: the expectation operator is denoted as $\mathbb{E}[\cdot]$, $[\cdot]^{T}$ denotes transpose, and $[\cdot]^{H}$ denotes conjugate transpose. $\Re\{\cdot\}$, $\Im\{\cdot\}$, $| \cdot |$, and $\text{arg}\{\cdot\}$ are the real, imaginary part, magnitude, and angle of a complex number, respectively. The Q-function is denoted as $\mathcal{Q}(\cdot)$.

\section{System Model}
\label{sec:system_model}

Let the received signal in the $k$th time slot be
\begin{IEEEeqnarray}{rcl}\label{eq:sig_mod_new}
{r}_{k} = x_{k}e^{j\theta_{k}} + n_{k},
\end{IEEEeqnarray}
where $x_{k}$ is the transmitted symbol, $\theta_{k}$ is the phase noise that is assumed Gaussian distributed with mean zero and variance $\sigma_\tr{p}^{2}$, i.e., $\theta_{k} \sim \mathcal{N}(0,\sigma_\tr{p}^{2})$, and ${n}_{k}$ is the complex Gaussian noise, i.e., ${n}_{k} \sim \mathcal{CN}(0,N_{0})$. Let ${\boldsymbol{r}} \triangleq \left[{r}_{0},\ldots,{r}_{L-1}\right]^{T}$ represent a vector of $L$ received symbols. The transmitted data is denoted as $\boldsymbol{x} \triangleq \left[x_{0},\ldots,x_{L-1}\right]^{T}$, where $x_{k}$ can assume any point in the signal constellation $\mathcal{X} = \{x^{(i)},\, \forall \, i \in \{1,\ldots,M\} \}$, and $M$ is the size of the constellation. Let {${\boldsymbol{\theta}}$}$ \triangleq \left[\theta_{0},\ldots,\theta_{L-1}\right]^{T}$ denote the phase noise vector. It is assumed that $\boldsymbol{x}$ and $\boldsymbol{\theta}$ are independent of each other. The vector ${\boldsymbol{n}}\triangleq \left[{n}_{0},\ldots,{n}_{L-1}\right]^{T}$ denotes a vector of independent identically distributed (i.i.d.) complex Gaussian random variables.

The vector $\boldsymbol{\theta}$ represents the residual phase error that results from tracking and compensating the phase of the received signal using an estimator. In general, $\boldsymbol{\theta}$ may be correlated and $\sigma_\tr{p}^{2}$ is a function of the received signals and the transmitted symbols and is known to the detector \cite{kam94}. However, in order to simplify the analysis, we assume that an ideal estimator is used that removes any correlation in $\boldsymbol{\theta}$. This assumption is widely used to analyze the performance of systems affected by phase noise \cite{vit63}, \cite{fos73}, \cite{imai04} \cite{kam09}, and also holds when the estimator operates in a data-aided mode \cite{marc99}. Thus, the sequence $\left[\theta_{0},\ldots,\theta_{L-1}\right]$ represents a \textit{memoryless phase noise process} \cite{imai04}.%Alternatively, we can assume that the sequence $\left[\theta_{0},\ldots,\theta_{L-1}\right]$ is memoryless to render our analysis independent of the phase estimation algorithm.

\section{ML Detection Methods}
\label{sec:analysis}

In this section, we first discuss the ML detector in the presence of phase noise as derived in \cite{kam94}. For the system model in \refeq{eq:sig_mod_new}, the likelihood based on the compensated received signal is
\begin{IEEEeqnarray}{rcl}\label{eq:mod_ml}
f(\boldsymbol{r}|x_{k}) &=& \int_{-\pi}^{\pi}p(r_{k}|x_{k},\theta_{k})p(\theta_{k}|\overline{\boldsymbol{r}}_{k},x_{k})\text{d}\theta_{k} ,
%&\triangleq& L_{i}(k), \nonumber
\end{IEEEeqnarray}
where $\overline{\boldsymbol{r}}_{k} \triangleq \left[r_{0},\ldots,r_{k-1},r_{k+1},\ldots,r_{L-1}\right]^{T}$ refers to all signals received outside the $k$th time instant, and the a posteriori probability density function (PDF) of the phase noise is  $p(\theta_{k}|\overline{\boldsymbol{r}}_{k},x_{k}) = p(\theta_{k}) = \mathcal{N}(0,\sigma_{\tr{p}}^{2})$. The ML decision for the $k$th symbol is given as
\begin{IEEEeqnarray}{rcl}\label{eq:dec_mod_ml}
\hat{x}_{k} =   \underset{x_{k}\in \mathcal{X}}{\operatorname{argmax}}\, f(\boldsymbol{r}|x_{k}).\nonumber
\end{IEEEeqnarray}
The likelihood function in \refeq{eq:mod_ml} is difficult to derive in its exact form. In the sequel we derive approximate forms of the likelihood function (and hence the approximate ML detectors) using assumptions on the SNR and phase noise. The approximate forms can be used to derive interesting performance metrics like SEP and MI that are associated with the detectors and the likelihood function.

\subsection{ML Detection Based on a High SNR Approximation}
We review an approximate ML detector that is derived using a high instantaneous SNR approximation in \cite{raj12}. Specifically, it is shown that the likelihood for the observation in the $k$th time instant (given $x_{k}$ is transmitted) can be written as
\begin{align}\label{eq:ml_gaussian_1}
f_{\text{SNR}}(r_{k}|x_{k})  &=&  \frac{e^{ -\frac{1}{2} \left(\frac{\left(\left|r_{k}\right| - \left|x_{k}\right|\right)^{2}}{N_{0}/2}  +
\frac{\left(\arg\{r_{k}\} - \arg\{x_{k}\}\right)^{2}}{\sigma_{\tr{p}}^{2} + \frac{N_{0}}{ 2{\left|x_{k}\right|}^{2}}} \right)}}{2\pi \sqrt{N_{0}/2 \left(|x_{k}|^{2}\sigma_{\tr{p}}^{2} + N_{0}/2\right)}},
\end{align}
where a high instantaneous SNR approximation is used, i.e., $|x_{k}| \gg \Re\{n_{k}\}$. Based on \refeq{eq:ml_gaussian_1}, the ML decision rule using the $k$th observation ($r_{k}$) is formulated as
\begin{IEEEeqnarray}{rCl}\label{eq:ml_gaussian}
\hat{x}_{k} &=&  \underset{x_{k}\in \mathcal{X}}{\operatorname{argmax}}\, f(\left| r_{k} \right|, \text{arg} \{ r_{k} \} |x_{k}) \nonumber\\
&=&  \underset{x_{k}\in \mathcal{X}}{\operatorname{argmin}}\, \frac{\left(\left|r_{k}\right| - \left|x_{k}\right|\right)^{2}}{N_{0}/2}  \nonumber\\
&+&  \frac{\left(\arg\{r_{k}\} - \arg\{x_{k}\}\right)^{2}}{\sigma_{\tr{p}}^{2} + \frac{N_{0}}{ 2{\left|x_{k}\right|}^{2}}} + \log\left(\sigma_{\tr{p}}^{2} \left|x_{k}\right|^{2} + {N_{0}/2}\right),\nonumber\\
&\triangleq&  \underset{x_{k} \in \mathcal{X}}{\operatorname{argmin}} \, L(x_{k}).
\end{IEEEeqnarray}

We refer to \refeq{eq:ml_gaussian} as the \emph{High SNR Gaussian PDF detector}, and denote it as \emph{GAP-D}. This detection rule and the likelihood function in \refeq{eq:ml_gaussian_1} are accurate for all phase noise scenarios and high instantaneous SNR (relatively low AWGN noise or high symbol energy).

Using \refeq{eq:ml_gaussian_1}, the MI of the memoryless phase noise channel \refeq{eq:sig_mod_new} can also be computed. However, due to the high SNR assumption, it may not be accurate for low SNRs.
\setcounter{equation}{13}
\begin{figure*}[!hb]
\HRule
\begin{IEEEeqnarray}{rcl}
    \label{eq:mean_var_eta}
a_{ij} \triangleq \frac{\left(\left|x^{(i)}\right| - \left|x^{(j)}\right|\right)^{2}}{N_{0}/2}, \;
b_{ij} &\triangleq& \frac{\left(\arg\{x^{(i)}\} - \arg\{x^{(j)}\}\right)^{2}}{\sigma_{\tr{p}}^{2} + \frac{N_{0}}{ 2{\left|x^{(j)}\right|}^{2}}},\;
c_{ij} \triangleq \frac{\sigma_{\tr{p}}^{2} + \frac{N_{0}}{ 2{\left|x^{(i)}\right|}^{2}}}{\sigma_{\tr{p}}^{2} + \frac{N_{0}}{ 2{\left|x^{(j)}\right|}^{2}}},\;
y_{ij} \triangleq  \log{\frac{\left({ \left|x^{(i)}\right|^{2}}\sigma_{\tr{p}}^{2} + \frac{N_{0}}{2}\right)}{\left({ \left|x^{(j)}\right|^{2}}\sigma_{\tr{p}}^{2} + \frac{N_{0}}{2}\right)}}\nonumber\\
\mathbb{E}\{\eta_{ij}|x^{(i)}\} &=&  1 - (a_{ij} + b_{ij} + c_{ij}),\nonumber\\
\mathbb{E}\left\{\left(\eta_{ij} - \mathbb{E}\{\eta_{ij}|x^{(i)}\}\right)^{2}|x^{(i)}\right\} &=& \mathbb{E}\{ L_{i}^{2}|x^{(i)}\} +  \mathbb{E}\{ L_{j}^{2} |x^{(i)}\} - 2\mathbb{E}\{ L_{i}L_{j} |x^{(i)}\} - \left(  \mathbb{E}\{\eta_{ij}|x^{(i)}\} \right)^{2}\nonumber\\
  &=& 2  + 4a_{ij} + 2c_{ij}^{2}  + 4b_{ij}c_{ij} - 4c_{ij}.
  \end{IEEEeqnarray}
\HRule
\end{figure*}

\subsection{ML Detection Based on a Low Instantaneous Phase Noise Approximation}
In the sequel, we derive a detector by using a low instantaneous phase noise approximation. Let the received signal model be as in \refeq{eq:sig_mod_new}, which after compensation in the $k$th time instant is written as
\setcounter{equation}{5}
\begin{subequations}
\begin{IEEEeqnarray}{rCl}\label{eq:sig_mod_new_2a}
r_{k}  &=& x_{k}e^{j\theta_{k}} + n_{k},\\\label{eq:sig_mod_new_2b}
&=& |x_{k}|e^{j\arg\{x_{k}\}}e^{j\theta_{k}} + n_{k}^{\prime}e^{j\arg\{x_{k}\}},\\\label{eq:sig_mod_new_2c}
&=& \left(|x_{k}|e^{j\theta_{k}} + n_{k}^{\prime}\right)e^{j\arg\{x_{k}\}},\\\label{eq:sig_mod_new_2d}
&\approx& \left(|x_{k}| + \Re \{n_{k}^{\prime}\} +  j(|x_{k}|\theta_{k} + \Im \{n_{ik}^{\prime} \} ) \right)e^{j\arg\{x_{k}\}},
\end{IEEEeqnarray}
\end{subequations}
where $n_{k}^{\prime} \triangleq n_{k}e^{-j\arg\{x_{k}\}}$ in \refeq{eq:sig_mod_new_2b}, and a low instantaneous phase noise approximation is used in \refeq{eq:sig_mod_new_2d} corresponding to $e^{j\theta_{k}} \approx 1 + j\theta_{k}$. Given $x_{k}$ and based on the real and imaginary parts of $r_{k}$ from \refeq{eq:sig_mod_new_2d}, we define
\begin{align}
u_{k} &\triangleq \Re \{r_{k}e^{-j\arg\{x_{k}\}}\} - |x_{k}|\nonumber\\
v_{k} &\triangleq \Im \{r_{k}e^{-j\arg\{x_{k}\}}\}.
\end{align}
The PDF of $[u_{k}, v_{k}]^{T}$ conditioned on $x_{k}$ is a bivariate Gaussian distribution with mean
\begin{align} \label{eq:mean_ml}
&\mathbb{E}[\Re \{r_{k}e^{-j\arg\{x_{k}\}} - |x_{k}|\}, \;  \Im \{r_{k}e^{-j\arg\{x_{k}\}}\} ] \nonumber\\
&= \mathbb{E}[ \Re \{n_{k}^{\prime}\}, \; |x_{k}|\theta_{k} + \Im \{n_{ik}^{\prime} \}  ] \nonumber\\
&= [0\, 0]^{T}.
\end{align}
The covariance of this conditional PDF is
\begin{IEEEeqnarray}{rcl} \label{eq:var_ml}
\mathbb{E}\left[
\begin{array}{cc}
 |\Re \{n_{k}^{\prime}\}|^{2} &  \Re \{n_{k}^{\prime}\}\left(|x_{k}|\theta_{k} + \Im \{n_{ik}^{\prime} \}\right) \\
\Re \{n_{k}^{\prime}\}\left(|x_{k}|\theta_{k} + \Im \{n_{ik}^{\prime} \}\right) & ||x_{k}|\theta_{k} + \Im \{n_{ik}^{\prime} \}|^{2}
\end{array}
\right] \nonumber\\
 = \left[
\begin{array}{cc}
N_{0}/2 & 0 \\
0 & \sigma_\tr{p}^{2}|x_{k}|^{2} + N_{0}/2
\end{array}
\right].\nonumber\\
\end{IEEEeqnarray}
Using  \refeq{eq:mean_ml} and  \refeq{eq:var_ml}, the likelihood of $r_{k}$, given $x_{k}$, based on the low instantaneous phase noise approximation is written as
\begin{align}\label{eq:ml_lowphn}
f_{\text{phn}} (r_{k}|x_{k})  &\triangleq f(u_{k}, v_{k}|x_{k}) \nonumber\\ & = \frac{e^{ -\frac{1}{2}\left(\frac{\left(\Re \{r_{k}e^{-j\arg\{x_{k}\}}\} - |x_{k}|\right)^{2}}{N_{0}/2}
+  \frac{\left(\Im \{r_{k}e^{-j\arg\{x_{k}\}}\}\right)^{2}}{\sigma_\tr{p}^{2}|x_{k}|^{2} + N_{0}/2  }\right)}}{2\pi\sqrt{N_{0}/2\left(\sigma_\tr{p}^{2}|x_{k}|^{2} + N_{0}/2 \right)}}.
%f_{2}(x_{k}) &\triangleq& \frac{\left(\Re \{r_{k}e^{-j\arg\{x_{k}\}}\} - |x_{k}|\right)^{2}}{N_{0}/2} \nonumber\\
%&+&  \frac{\left(\Im \{r_{k}e^{-j\arg\{x_{k}\}}\}\right)^{2}}{\sigma_\tr{p}^{2}|x_{k}|^{2} + N_{0}/2  }.\nonumber
\end{align}
Using this likelihood function, an ML decision rule can be derived based on the $k$th observation $r_{k}$  as
\begin{IEEEeqnarray}{rCl}\label{eq:ml_lpn}
\hat{x}_{k} &=&  \underset{x_{k}\in \mathcal{X}}{\operatorname{argmax}}\, f( r_{k}  |x_{k}) \nonumber\\
&=&  \underset{x_{k}\in \mathcal{X}}{\operatorname{argmin}}\, \frac{\left(\Re \{r_{k}e^{-j\arg\{x_{k}\}}\} - |x_{k}|\right)^{2}}{N_{0}/2} \nonumber\\
&+&  \frac{\left(\Im \{r_{k}e^{-j\arg\{x_{k}\}}\}\right)^{2}}{\sigma_\tr{p}^{2}|x_{k}|^{2} + N_{0}/2  } + \log \left( {\sigma_\tr{p}^{2}|x_{k}|^{2} + N_{0}/2  }\right).\nonumber\\
\end{IEEEeqnarray}
We call the detector in \refeq{eq:ml_lpn} the \emph{Low Phase Noise Detector} and denote it as \emph{LPN-D}. This detection rule and the likelihood function in \refeq{eq:ml_lowphn} are accurate for all SNR values and practically high phase noise variance (i.e., values up to around $\sigma_{\tr{p}}^{2} = 0.01 \text{ rad}^{2} $ \cite{moen09}).

The MI of the memoryless phase noise channel can also be computed based on \refeq{eq:ml_lowphn}. However, it may not be accurate for extremely high phase noise variance ($\sigma_{\tr{p}}^{2} > 0.1 \text{ rad}^{2}$).

\section{Performance Metrics}
\label{sec:metrics}

In this section, we discuss interesting performance metrics associated with the detectors and the likelihood functions derived in Section \ref{sec:analysis}.

\subsection{SEP for GAP-D}

We review the SEP derived for GAP-D in \cite{raj12}. The SEP for GAP-D is derived by averaging over all pairwise symbol error probabilities (union bound) \cite{proakis}
\begin{IEEEeqnarray}{rcl}\label{eq:error_bound}
P_{\tr e} \leq \frac{1}{M} \sum_{i=1}^{M} \sum_{j = 1, j \neq i }^{M}\text{Pr} \left( L_{i} - L_{j} > 0 | x^{(i)} \right),
\end{IEEEeqnarray}
where $L_i \triangleq L(x^{(i)})$ for $x^{(i)} \in \mathcal{X}$, and $\text{Pr} \left( L_{i} - L_{j} > 0 | x^{(i)} \right)$ is the probability of a pairwise symbol error event expressed in terms of the GAP-D metrics that are defined implicitly in \refeq{eq:ml_gaussian}. This event corresponds to the case when the transmitted symbol is detected as $x^{(j)} \in \mathcal{X}$, given ${x^{(i)} \in \mathcal{X}, i \neq j}$, is transmitted. Define $\eta_{ij} \triangleq L_{i} -  L_{j}$. Then, as in \cite{raj12}, a high SNR approximation can be applied to simplify $\text{Pr} \left( L_{i} - L_{j} > 0 | x^{(i)} \right)$ $= \text{Pr} \left( \eta_{ij} > 0 | x^{(i)} \right) $ as
\begin{IEEEeqnarray}{rcl}\label{eq:p_error_etaij}
\text{Pr}\left(\eta_{ij} > 0 | x^{(i)}\right) \approx \mathcal{Q}  \left( \frac{y_{ij} - \mathbb{E}\{\eta_{ij}|x^{(i)}\} }{\sqrt{  \mathbb{E}\left\{\left(\eta_{ij} - \mathbb{E}\{\eta_{ij}|x^{(i)}\}\right)^{2}|x^{(i)}\right\}  }} \right),\nonumber\\
\end{IEEEeqnarray}
where the terms in the argument are defined in \refeq{eq:mean_var_eta}. Using \refeq{eq:p_error_etaij} in \refeq{eq:error_bound}, the probability of error for GAP-D is upper bounded as
\setcounter{equation}{14}
\begin{IEEEeqnarray}{rcl}\label{eq:p_error}
P_{\tr e} &\leq& \frac{1}{M}\sum_{i = 1}^{M}\sum_{j = 1,j \neq i}^{M}  \mathcal{Q}  \left( \frac{y_{ij} - \mathbb{E}\{\eta_{ij}|x^{(i)}\} }{\sqrt{  \mathbb{E}\left\{\left(\eta_{ij} - \mathbb{E}\{\eta_{ij}|x^{(i)}\}\right)^{2}|x^{(i)}\right\}  }} \right), \nonumber\\
&\triangleq&\, P_{\tr{error}}(\mathcal{X}).
\end{IEEEeqnarray}
This is the SEP target function that will be minimized to design constellations in the first formulation.

\subsubsection{SEP at High SNR and Error Floors}
The SEP of a given constellation at asymptotically high SNR (or error floor) can be obtained by evaluating $\underset{N_{0} \to 0} \lim \text{Pr}\left(\eta_{ij} > 0 | x^{(i)}\right)$  \cite{raj12} from \refeq{eq:mean_var_eta} and \refeq{eq:p_error} as
\begin{IEEEeqnarray}{rcl} \label{eq:error_floor}
\underset{N_{0} \to 0}\lim  \text{Pr}\left(\eta_{ij} > 0 | x^{(i)}\right)  = \mathcal{Q} \left( \sqrt{\frac{\arg\{x^{(j)}\} - \arg\{x^{(i)}\}}{\sigma_{\tr{p}}  }  } \right).\nonumber\\
\end{IEEEeqnarray}
Upon applying the union bound in \refeq{eq:error_bound} by considering only those pairs of symbols with equal energy, the error floor can be expressed as
\begin{align}
\underset{N_{0} \to 0}\lim P_{\tr e} \leq \frac{1}{M}\sum_{i = 1}^{M}\sum_{\substack { j = 1, j \neq i,\\ |x^{(i)}|=|x^{(j)}|}}^{M} \mathcal{Q} \left( \sqrt{\frac{\arg\{x^{(j)}\} - \arg\{x^{(i)}\}}{\sigma_{\tr{p}}  }  } \right).
\end{align}

For details about the derivation of the results in \refeq{eq:p_error}, \refeq{eq:error_floor}, we refer the reader to \cite{raj12}.

\subsection{MI of the Discrete Channel with Memoryless Phase Noise, AWGN and GAP-D}

We derive the MI of the discrete channel consisting of memoryless phase noise, AWGN, and the GAP-D ML detector. Let $X$ denote the input (transmitted symbol) to the effective channel, and $\hat{X}$ denote the decision made by GAP-D, where $X, \hat{X}  \in \mathcal{X}$. Then, given that all symbol points are equally likely, the MI \cite{cover} of this discrete-input  discrete-output channel  is
\begin{IEEEeqnarray}{rCl} \label{eq:mi_mldet1}
I_{\text{DD}}(X;\hat{X}) &=& H(X) - H(X|\hat{X})\nonumber\\
%&=& \log_{2}M + \sum P(X,\hat{X}) \log_{2}( P(X|\hat{X}) ),\nonumber\\
%&=& \log_{2}M +  P(\hat{X}|X)P(X) \log_{2}\left(\frac{ P(\hat{X}| x^{(i)})P(x^{(i)}) }{P(\hat{X}) } \right),\nonumber\\
&=& \log_{2}M +  \sum_{j=1}^{M}\sum_{i=1}^{M} P({\hat{x}^{(j)}}|x^{(i)})P(x^{(i)}) \nonumber\\
&&\log_{2}\left(\frac{ P({\hat{x}^{(j)}}| x^{(i)}) }{ \sum_{i=1}^{M} P({\hat{x}^{(j)}}|x^{(i)})} \right),
\end{IEEEeqnarray}
where $P({\hat{x}^{(j)}}|x^{(i)})$ is the probability that symbol ${\hat{x}^{(j)}} \in \mathcal{X}$ is detected when symbol $x^{(i)} \in \mathcal{X}$ is transmitted.

Here, we consider two cases. For the first case, let ${\hat{x}^{(j)}} \neq x^{(i)}$ . Then
\begin{IEEEeqnarray}{rCl}\label{eq:mi_mldet2}
P({\hat{x}^{(j)}}|x^{(i)}) &=& \text{Pr}(L_{j} < \underset{k \neq j}{\operatorname{min}} \,L_{k} | x^{(i)})\nonumber\\
&=& \text{Pr}( L_{i} - L_{j}  > 0|x^{(i)}) \prod_{\substack{k = 1,\\k \neq i,j}}^{M}\nonumber\\
&& \text{Pr}( L_{k} - L_{j} > 0|x^{(i)},L_{j} < L_{k-1},L_{j} < L_{i} ),\nonumber\\
&\approx& \text{Pr}( L_{i} - L_{j}  > 0|x^{(i)}),
\end{IEEEeqnarray}
where \refeq{eq:mi_mldet2} is given in \refeq{eq:p_error_etaij} and \refeq{eq:mean_var_eta}, and $L_{i},L_{j}$ and $L_{k}$ are the GAP-D metrics defined implicitly in \refeq{eq:ml_gaussian}. The conditional probability $P({\hat{x}^{(j)}}|x^{(i)})$ can be approximated by using $\text{Pr}( L_{k} - L_{j}  > 0 |x^{(i)},L_{j} < L_{k-1},L_{j} < L_{i} ) \approx 1$. That is, it is assumed that $\text{Pr}(L_{k} - L_{j} > 0) = 1$ for $k \neq i,j$, given $L_{j} < L_{i}$ and $x^{(i)}$ is the transmitted symbol. It can be verified by simulations that this approximation is tight for medium-to-high SNR scenarios.

In the second case, let  ${\hat{x}^{(j)}} = x^{(i)}$. Then
\begin{IEEEeqnarray}{rCl}\label{eq:mi_mldet4}
P({\hat{x}^{(j)}}|x^{(i)}) \approx   1 - \sum_{\substack{j = 1,\\j \neq i}}^{M} P({\hat{x}^{(j)}}|x^{(i)}).
\end{IEEEeqnarray}
Plugging \refeq{eq:mi_mldet2} and \refeq{eq:mi_mldet4} in \refeq{eq:mi_mldet1} gives the approximate MI of the desired channel.

\subsection{MI of the Memoryless Phase Noise Channel}
\begin{figure}[!t]
\begin{center}
\includegraphics[width = 3.5in, keepaspectratio=true]{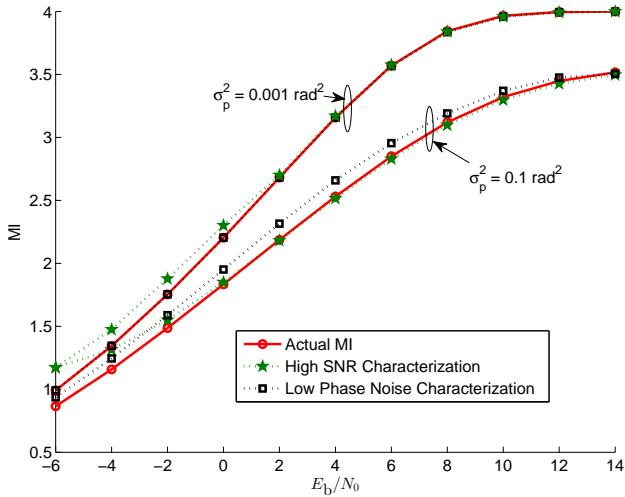}
\caption{ MI for $16$-QAM for $\sigma_\tr{p}^{2} = 0.1, 0.001\, \text{rad}^{2}$ and different $E_{\tr b}/N_{0}$   }\label{fig:mi_ana}
\end{center}
\end{figure}

We consider the memoryless phase noise channel in \refeq{eq:sig_mod_new}, where the received signal is compensated with an ideal estimator, and we characterize the MI of this discrete-input continuous-output channel. Note that this channel is different from the discrete-input discrete-output channel that was considered previously. Let $X \in \mathcal{X}$ denote the transmitted symbols, and let $R \in \mathbb{C}^{2}$ denote the received signal after compensation (dropping time index $k$). Given that all symbol points are equally likely, the MI for this channel with discrete-input ($X$) and continuous-output ($R$) is
\begin{IEEEeqnarray}{rCl}\label{eq:mi_mlmemoryless}
I_{\text{DC}}(X;R) &=& H(X) - H(X|R)\nonumber\\
%&=& \log_{2}M + \sum P(X,\hat{X}) \log_{2}( P(X|\hat{X}) ),\nonumber\\
%&=& \log_{2}M +  P(\hat{X}|X)P(X) \log_{2}\left(\frac{ P(\hat{X}| x^{(i)})P(x^{(i)}) }{P(\hat{X}) } \right),\nonumber\\
&=& \log_{2}M +  \sum_{i=1}^{M}\int f(r|x^{(i)})P(x^{(i)}) \nonumber\\
&\times&\log_{2}\left(\frac{ f(r| x^{(i)}) }{ \sum_{i=1}^{M} f(r|x^{(i)})} \right) \text{d} r.
%&=& \log_{2}M +  \sum_{i=1}^{M}\int_{r} f(r|x^{(i)})P(x^{(i)}) \log_{2}\left(\frac{ f(r| x^{(i)}) }{ \sum_{i=1}^{M} f(r|x^{(i)})} \right).
\end{IEEEeqnarray}
The MI of this channel is evaluated by using two different characterizations. In the first characterization, it is computed using the likelihood function $f_{\text{SNR}}(r| x^{(i)})$ based on the high SNR approximation in \refeq{eq:ml_gaussian_1}. For the second characterization, we use the likelihood function  $f_{\text{phn}}(r| x^{(i)})$ in \refeq{eq:ml_lowphn}, which is based on the low instantaneous phase noise approximation. Based on these likelihoods, the MI can now be accurately evaluated for the following scenarios: (a) medium-to-high SNR, any phase noise variance; (b) any SNR, low-to-medium phase noise variance. The only scenario where these characterizations do not render an accurate MI is when the SNR is low and the phase noise variance is high. The accuracy of these characterizations is demonstrated for the $16$-QAM constellation in Fig. \ref{fig:mi_ana}. As expected, at low SNR the MI based on the high SNR approximation is relatively inaccurate, while for high phase noise variance ($\sigma_\tr{p}^{2} = 0.1$ rad$^{2}$), the MI based on the instantaneous low phase noise approximation is more inaccurate. Compared to the MI derived for a similar channel in \cite{tor12}, our characterizations are analytically simpler and are accurate for a wide range of SNR values and phase noise variances.

\section{Constellation Optimization}
\label{sec:cons_opt}

In this section, we present three optimization formulations based on $P_{\tr{e}}(\mathcal{X})$, $I_{\text{DD}}(X;\hat{X})$ and $I_{\text{DC}}(X;R)$  to design constellations of order $M$, and we adopt a global optimization approach to solve them. Then, for each performance criterion, we present a detailed analysis of the optimal placement of the symbol points.

\subsection{Optimization Formulations}

In the first formulation, we seek to design constellations that minimize the SEP of GAP-D for a fixed $E_{\tr b}/N_{0}$ (SNR per bit), and phase noise variance $\sigma_{\tr{p}}^{2}$. The optimization problem is posed as follows.
\begin{align}\label{eq:sep_opt_form}
& \underset{{  \mathcal{X}} }{\operatorname{minimize}} \; P_{\tr{e}}(\mathcal{X}),
\nonumber\\
&{\operatorname{subject}} \, {\operatorname{to}} \; \frac{1}{M} \sum_{i=1}^{M} x^{(i)}{x^{(i)}}^{*} \leq P.
\end{align}
For future reference, we refer to this problem as the SEP formulation and denote it as SEP-A.

In the next formulation, we determine optimal constellations that maximize the MI of the discrete channel consisting of memoryless phase noise, AWGN and the GAP-D ML detector as
\begin{align}\label{eq:mia_opt_form}
&\underset{{  \mathcal{X}} }{\operatorname{maximize}} \; I_{\text{DD}}(X;\hat{X}),
\nonumber\\
&{\operatorname{subject}} \, {\operatorname{to}} \; \frac{1}{M} \sum_{i=1}^{M} x^{(i)}{x^{(i)}}^{*} \leq P.
\end{align}
We refer to this problem as the MI formulation for GAP-D, and denote it as MI-A.

In the third formulation, we seek to determine optimal constellations of order $M$ that maximize the MI of the memoryless phase noise channel \refeq{eq:sig_mod_new} as
\begin{align}\label{eq:mib_opt_form}
& \underset{{ \mathcal{X}} }{\operatorname{maximize}} \; I_{\text{DC}}(X;R),
\nonumber\\
&{\operatorname{subject}} \, {\operatorname{to}} \; \frac{1}{M} \sum_{i=1}^{M} x^{(i)}{x^{(i)}}^{*} \leq P.
\end{align}
The MI for the phase noise channel is computed by using both the likelihoods $f_{\text{SNR}}(r|x^{(i)})$ and $f_{\text{phn}}(r|x^{(i)})$  in \refeq{eq:mi_mlmemoryless}. Using both characterizations of the MI as the objective function, two sets of constellations are obtained for different values of $\sigma_\tr{p}^{2}$ and $E_{\tr b}/N_{0}$. Then, for a given value of $E_{\tr b}/N_{0}$ and $\sigma_\tr{p}^{2}$, we numerically compute the actual MI \refeq{eq:mi_mlmemoryless} of the two constellations, and the constellation with the higher MI is picked as optimal. Note that the evaluation of the MI in \refeq{eq:mi_mlmemoryless} involves a double integral, which is computed using a composite trapezoidal rule \cite{trapz}. This optimization problem is referred to as the MI formulation and denoted as MI-B.

For the formulations in \refeq{eq:sep_opt_form}, \refeq{eq:mia_opt_form} and \refeq{eq:mib_opt_form}, $P$ denotes an average power constraint. The symbols denoted as $x^{(i)}$ are the optimization variables, which are continuous and belong to the complex plane. The assumption that all symbols $x^{(i)}$ are equally likely is implicit here. All formulations considered are non-linear optimization problems and are non-convex in $x^{(i)}$. Hence, the solutions obtained are not guaranteed to be globally optimal. We solve the optimization problems by a numerical global search method as in \cite{ugray07}, which is implemented using the MATLAB Global Optimization Toolbox. This method is a gradient-based algorithm that uses multiple randomized starting points to find and compare different local optima of a smooth nonlinear optimization problem.

\subsection{Results and Discussion}
We consider different values of $E_{\tr b}/N_{0}$ (-$2$--$20$ dB), $\sigma_\tr{p}^{2} = 0.01$ and $0.1$ $\text{rad}^{2}$, and $M =16$. The chosen values of $\sigma_\tr{p}^{2}$ correspond to the variance of the phase estimator in strong phase noise scenarios \cite{moen09}. Note that constellations for other values of $M$ can be obtained by the same global search method. However, the complexity of the optimization algorithm scales exponentially with $M$. {In general, the optimal constellations obtained from all the formulations are non-symmetric. We conjecture that this is because of the objective functions and the power constraint used. This claim is supported by the fact that (symmetric) APSK constellations achieve lower MI.}

\subsubsection{Optimal SEP-A Constellations}
\begin{figure}[!t]
\begin{center}
\includegraphics[height = 3.2in, width = 3.6in ]{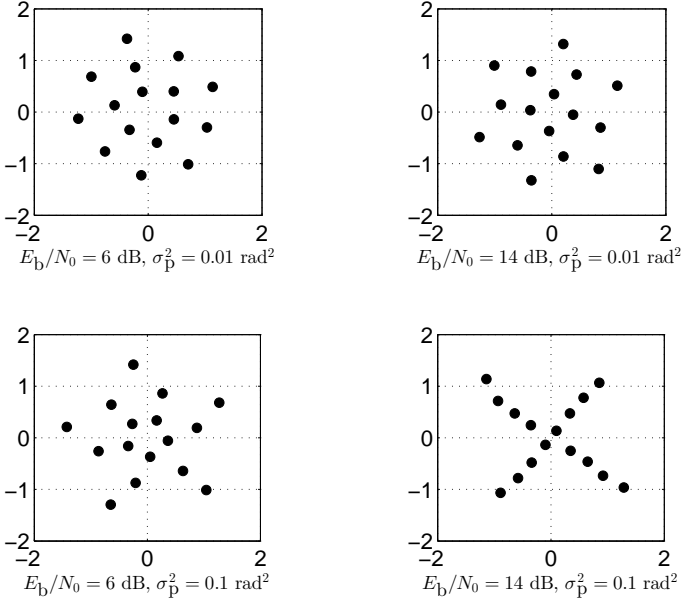}
\caption{ Constellations optimizing SEP-A for $\sigma_\tr{p}^{2} = 0.1, 0.01\, \text{rad}^{2}$ and different $E_{\tr b}/N_{0}$  }\label{fig:sep_const}
\end{center}
\end{figure}

The optimal SEP-A constellations obtained for different values of $E_{\tr b}/N_{0}$ and $\sigma_\tr{p}^{2}$ are given in Fig. \ref{fig:sep_const}, and we draw the following conclusions.
\begin{itemize}[leftmargin=*]
\item For a fixed $\sigma_\tr{p}^{2}$, the number of energy levels in the optimal constellations gradually increases with increase in $E_{\tr b}/N_{0}$. This is similar to the observation in \cite{chris12}.
\item As $\sigma_\tr{p}^{2}$ is increased for a fixed $E_{\tr b}/N_{0}$, the number of energy levels in the optimal constellations increases.
\item For all values of $E_{\tr b}/N_{0}$ and $\sigma_\tr{p}^{2}$, symbol points that are of the same energy are separated by the largest possible angular distance.
\end{itemize}

The union bound on the SEP derived in \refeq{eq:p_error} is inaccurate for low SNR, rendering the optimization formulation (SEP-A) sub-optimal in such scenarios. However, as we shall see in Section \ref{sec:results}, constellations optimized using SEP-A have similar SEP at low SNR, and much better performance at high SNR, than all other constellations proposed in prior work. %%Constellations with a symbol point at the origin are known to perform well in phase noise scenarios \cite{fos73}.

%\HRule
%\HRule

\subsubsection{Optimal MI-A Constellations}

For a given $E_{\tr b}/N_{0}$ and phase noise variance, we observe that the constellations that are optimal for MI-A are also those that optimize SEP-A. Note that the MI derived for the effective channel is a function of the pairwise error probability of the symbols as in \refeq{eq:mi_mldet1}. Even though maximizing the MI of the effective channel is equivalent to minimizing a function of the pairwise error probability, the optimal solutions for MI-A and SEP-A may not be the same since the optimization formulations are different and cannot be verified to have the same global optima. The MI of optimal constellations for this formulation is presented in Section \ref{sec:results}.

%\label{sec:cons_opt_mib}
%\HRule
\begin{figure}[t]
\begin{center}
\includegraphics[height = 3.3in, width = 3.5in]{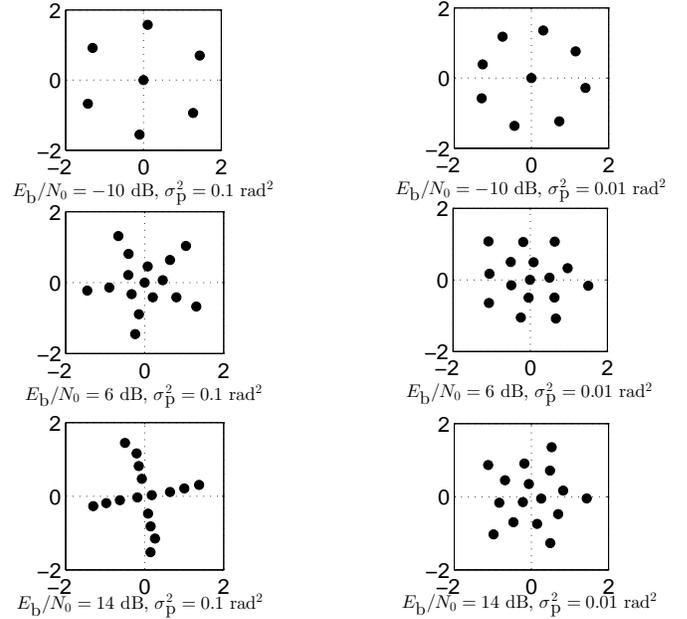}
\caption{ Constellations optimizing MI-B for $\sigma_\tr{p}^{2} = 0.1, 0.01 \, \text{rad}^{2}$ and different $E_{\tr b}/N_{0}$   }\label{fig:mib_const}
\end{center}
\end{figure}
%\HRule

\subsubsection{Optimal MI-B Constellations}

The optimal MI-B constellations are presented in Fig. \ref{fig:mib_const} for different values of $E_{\tr b}/N_{0}$ and $\sigma_\tr{p}^{2}$. From Fig. \ref{fig:mib_const} and our simulations, we make the following observations.
\begin{itemize}[leftmargin=*]
\item At low $E_{\tr b}/N_{0}$, the optimal constellations obtained using the MI based on $f_{\text{phn}}(r|x^{(i)})$  have higher MI than those obtained from the characterization based on $f_{\text{SNR}}(r|x^{(i)})$. This is because the likelihood based on $f_{\text{SNR}}(r|x^{(i)})$ is more inaccurate in this scenario.
\item As $E_{\tr b}/N_{0}$ is increased, constellations obtained by optimizing the MI based on $f_{\text{SNR}}(r|x^{(i)})$ achieve higher MI than those obtained from the  characterization based on $f_{\text{phn}}(r|x^{(i)})$.
\item At low $E_{\tr b}/N_{0}$, we observe that the optimal constellations are similar to those obtained for an AWGN dominated channel \cite{fos74}. Also, a point is observed to occur at the origin for low-to-medium values of $E_{\tr b}/N_{0}$, similar to the observation in \cite{katz04}.
\item {For very low values of $E_{\tr b}/N_{0}$, the number of symbol points in the optimized constellations is observed to be less than $M=16$. In particular, some points are superimposed on the point at the origin, which suggests probabilistic shaping for $M=16$. However, we performed a similar optimization with a reduced number of points ($M < 16$) obtaining a similar MI. This indicates that symbol points above a certain number do not contribute much to the MI. This observation is in line with \cite[Theorem 2]{colavolpe11}.}
    \item As $E_{\tr b}/N_{0}$ increases, the number of energy levels of the symbol points in the optimized constellation increases, and the  point at the origin disappears. Furthermore, the symbol points of the same energy are separated by the largest possible angular distance.
\item As the phase noise variance is increased for a fixed $E_{\tr b}/N_{0}$, the number of energy levels in the optimized constellations increases.
\end{itemize}

%\subsubsection{Joint Optimization of Probability Distribution of Symbols and their Geometric Arrangement}
%
%In formulation MI-B (\textcolor{red}{actually for all formulations. I would say at the beginning that, unless otherwise stated, we consider equally-likely symbols. There is too much redundancy in the paper right now.}), it is assumed that all the $M$ symbols in the constellation are equally likely, i.e., $P(x^{(i)})=1/M, \, x^{(i)} \in \mathcal{X}$ (\textcolor{red}{move this at the beginning}). It is possible to remove the uniform distribution assumption on the symbols, and jointly optimize the probability distribution $P(x^{(i)})$ and their geometric arrangement such that the MI is maximized. This procedure can be summarized as follows.
%\begin{enumerate}
%\item Initialize $P(x^{(i)})$ and $f(r|x^{(i)})$, $x^{(i)} \in \mathcal{X}$.
%\item For a given $P(x^{(i)})$, determine the optimum values of $P(r|x^{(i)})$.
%\item For a given $P(r|x^{(i)})$, determine the optimum values of $P(x^{(i)})$. Given $P(r|x^{(i)})$, MI is concave in $P(x^{(i)})$ \cite{cover}.
%\item Execute iteratively steps 1) and 2) until convergence.
%\end{enumerate}
%The MI obtained from the procedure above renders an upper bound on the MI or the constrained capacity \textcolor{red}{(???)} of the channel when the input set cardinality ($|\mathcal{X}|$) is constrained to a finite value $M$. We refer to the MI obtained from this algorithm as the MI bound for a given $M$.

\section{Design of Structured Constellations}
\label{sec:ring_const}

\begin{figure}[!t]
\begin{center}
\includegraphics[width = 3.4in, height = 2.4in, keepaspectratio=true]{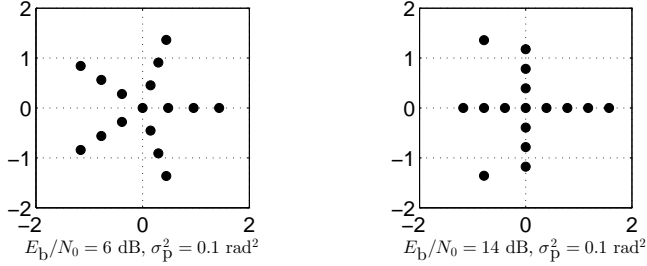}
\caption{ Optimized APSK constellations for  $\sigma_{\tr{p}}^{2} = 0.1 \text{ rad}^{2}$ (a) $E_{\tr b}/N_{0} = 6$ dB, (b) $E_{\tr b}/N_{0} = 14$ dB.}\label{fig:APSK_06_14}
\end{center}
\end{figure}

The global optimization approach adopted in formulations SEP-A, MI-A and MI-B results in constellations that depend on $E_{\tr b}/N_{0}$ and $\sigma_{\tr{p}}^{2}$, and are unstructured, i.e., they do not possess well-defined structures like, e.g., APSK constellations \cite{chris12} or spiral-shaped constellations \cite{kwa08}. Constraining the constellation set to a particular structure has several advantages. {The transceiver design becomes more practical as the APSK configurations can be defined  by fewer quantization levels for the amplitude and phase of the transmitted symbols and they have simpler decoding regions \cite{sam98}.} Also, the set of optimized structured constellations for a wide range of $E_{\tr b}/N_{0}$ and $\sigma_{\tr{p}}^{2}$ values is typically smaller, which can be helpful when performing adaptive modulation and coding. Finally, issues related to the local minima can be circumvented by means of an exhaustive search over a smaller (finite) set of constellation configurations.

{In this section, we construct simple APSK constellations as in \cite{chris12,chris13} that optimize the SEP and MI criterion used in SEP-A, MI-A, and MI-B. We use the routine in \cite{chris12,chris13} that optimizes the total number of rings in an APSK configuration and the number of points $n_{l}$, $1 \leq n_{l} \leq M$, distributed in its $l$th ring ($1\leq l \leq M$). Let $r_{l}$ denote the radius of the $l$th ring and $\phi_{l}$ denote the phase offset of all the points in it. Furthermore, we denote the $i$th APSK configuration as $A^{(i)}$, and the set of all APSK configurations as $\mathcal{A}$. The ring configuration $A^{(i)} \in \mathcal{A}$ is defined as
\begin{IEEEeqnarray}{rCl} \label{eq:apsk}
A^{(i)} = \left\{ r_{l}e^{j\left( \frac{2 \pi j}{n_{l}} + \phi_{l}\right)} : 1 \leq l \leq M, 0 \leq j \leq  n_{l} - 1  \right\}
\end{IEEEeqnarray}
In order to reduce the optimization search space, the ring radii are assumed to be ordered and uniformly spaced, i.e., $r_{1}(A^{(i)}) < \ldots < r_{M}(A^{(i)})$ and $r_{l+1} - r_{l} = \delta$. In addition, the $l$th ring is constrained to have a zero phase offset $\phi_{l} = 0$. For specifying the structured ring constellation $A^{(i)}$, we define $\boldsymbol{n}(A^{(i)}) = (n_{1}(A^{(i)}),\ldots,n_{{l}}(A^{(i)}))$, $\boldsymbol{r}(A^{(i)}) = (r_{1}(A^{(i)}),\ldots,r_{{l}}(A^{(i)}))$ and $\boldsymbol{\phi}(A^{(i)}) = (\phi_{1}(A^{(i)}),\ldots,\phi_{l}(A^{(i)}))$. Furthermore, for $n_{1}(A^{(i)})= 1$ , the point in the first ring is always placed at the origin, implying $r_{1} = 0$. If $n_{1}(A^{(i)}) \geq 2$, then $r_{1}(A^{(i)}) = \delta$. For brevity, when specifying the ring configuration, we use the notation $\boldsymbol{n}$-APSK, thereby omitting the radii and phase offset information.}

The optimization routine evaluates the desired criterion (SEP and MI) for all configurations and determines the optimized APSK configuration exhaustively. Formally, we define this optimization problem as
\begin{IEEEeqnarray}{rcl} \label{eq:ring_apsk}
&& \underset{A^{(i)} \in \mathcal{A}}{\operatorname{minimize}} \; \{P_{\tr{e}}(A^{(i)}), -I_{\text{DD}}(A^{(i)}),-I_{\text{DC}}(A^{(i)})\}\\
&&{\operatorname{subject}} \, {\operatorname{to}} \; \frac{1}{M} \sum_{n_{l}(A^{(i)}) \in \mathcal{A}} n_{l}(A^{(i)})|r_{n_{l}(A^{(i)})}|^{2} \leq P,
\nonumber
\end{IEEEeqnarray}
where, for brevity, the criterion written as $P_{\tr{e}}(A^{(i)}), -I_{\text{DD}}(A^{(i)}),-I_{\text{DC}}(A^{(i)})$ denotes the SEP and MI of the APSK configuration $A^{(i)}$. Note that when $I_{\text{DC}}(A^{(i)})$ is used as the objective function, the optimization technique used is similar to that used in formulation MI-B. For future reference, we refer to the general optimization formulation in \refeq{eq:ring_apsk} as the APSK formulation and denote it as APSK-A.

{For demonstrative purposes, in Fig. \ref{fig:APSK_06_14}, we present the optimized ring constellations obtained by using $I_{\text{DC}}(A^{(i)})$ as the objective function in formulation APSK-A. For $\sigma_{\tr{p}}^{2} = 0.1 \text{ rad}^{2}$, $E_{\tr b}/N_{0}= 6$ dB and $E_{\tr b}/N_{0}= 14$ dB, the optimized APSK configurations are (1,5,5,5)-APSK and (1,4,4,4,3)-APSK constellations respectively. Comparing these optimized APSK constellations with the MI-B optimizing constellation from Fig. \ref{fig:mib_const} for similar $\sigma_{\tr{p}}^{2}$ and  $E_{\tr b}/N_{0}$ values, we observe that they are very similar in the number of rings and the points distribution per ring.}

{Note that the number of rings, the ring radii, the number of points per ring and the phase offsets of APSK constellations can also be optimized using a more complex routine where we first pre-define the number of rings, and then determine the optimal number of points per ring, radii and phase offset for each ring. We do not present this routine in this paper because the resulting constellations perform only marginally better than those obtained from the APSK-A formulation in \refeq{eq:ring_apsk}. Also, as we shall see in Sec. \ref{sec:results}, the performances of the constellations obtained from APSK-A are close to that of those obtained from the global optimization routines. This implies that the routine used in (26) suffices to find good APSK constellations. A similar observation is also made in \cite{chris13}.}

\begin{figure*}[!t]
\begin{center}
\begin{tabular}{cc}
\includegraphics[width = 3.5in, keepaspectratio=true]{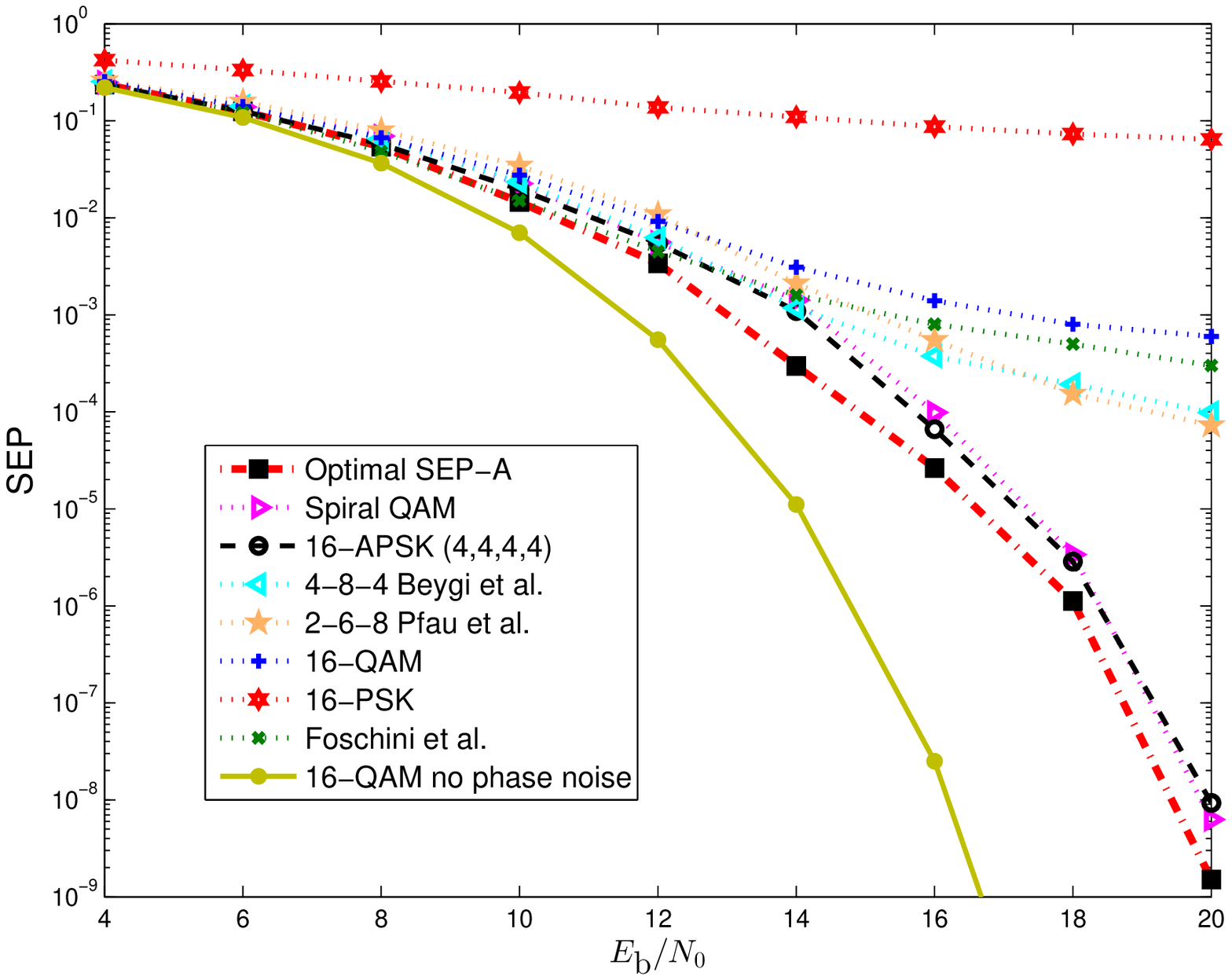}
&
\includegraphics[width = 3.65in, keepaspectratio=true]{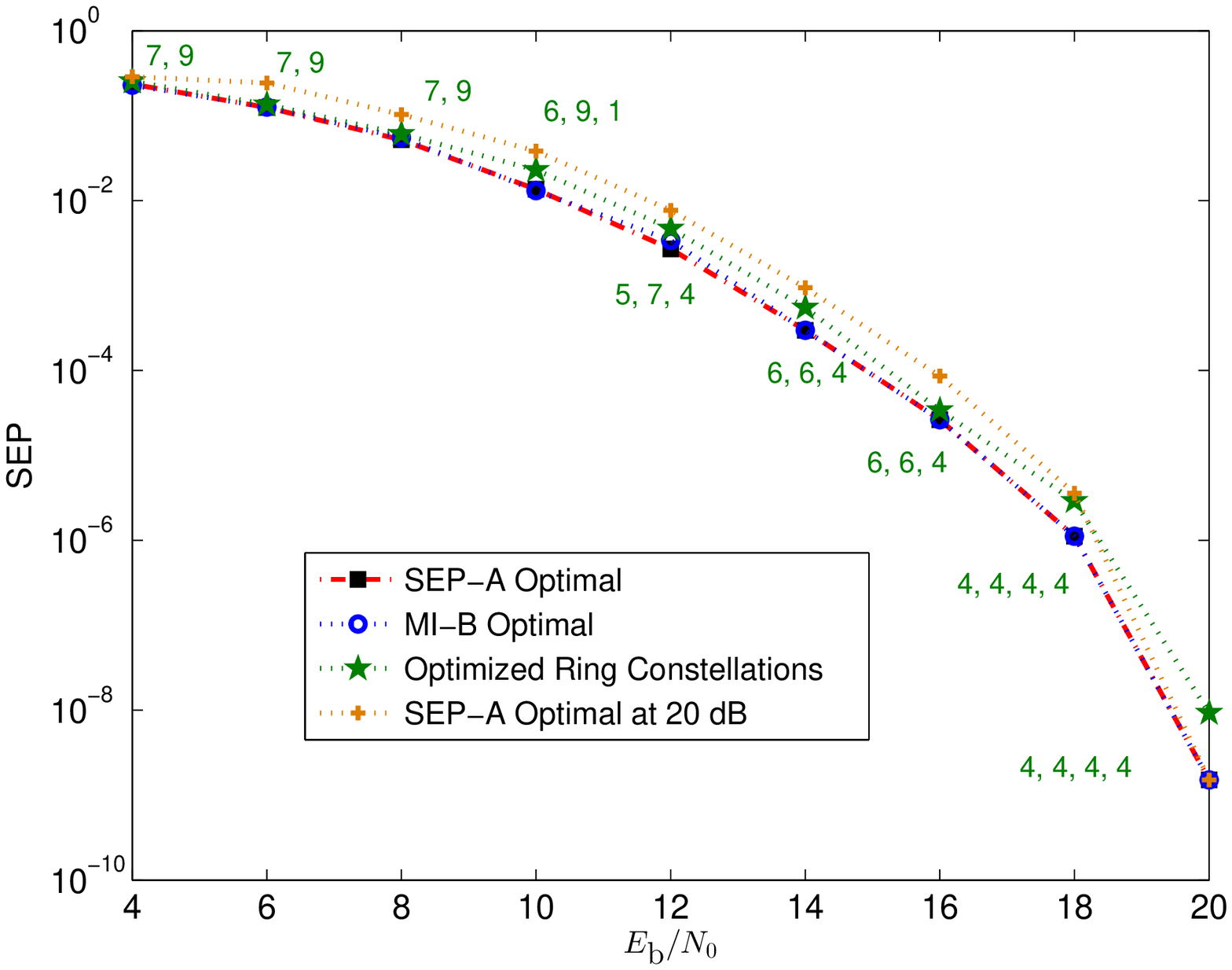}
\\
(a) & (b)\\
\end{tabular}
\caption{(a) SEP for different constellations for $\sigma_\tr{p}^{2} = 0.01\, \text{rad}^{2}$. (b) SEP for SEP-A, MI-B optimal constellations and the optimal APSK constellations (along with the optimal ring configurations) obtained by optimizing $P_{\tr{e}}(A^{(i)})$.}\label{fig:sep_comp}
\end{center}
\end{figure*}
%\begin{figure*}[!ht]
%\begin{center}
%\begin{tabular}{cc}
%\includegraphics[width = 3.5in, keepaspectratio=true]{APSK6_01.eps}
%&
%\includegraphics[width = 3.5in, keepaspectratio=true]{APSK14_01.eps}
%\\
%(a) & (b)\\
%\end{tabular}
%\caption{APSK constellations obtained by optimizing $I_{\text{DC}}(A^{(i)})$ for  $\sigma_{\tr{p}}^{2} = 0.1 \text{ rad}^{2}$ (a) $E_{\tr b}/N_{0} = 6$ dB, (a) $E_{\tr b}/N_{0} = 14$ dB.}\label{fig:APSK_0614_01}
%\end{center}
%\end{figure*}

\section{Comparative Study}
\label{sec:results}
In this section, we study the performance of the optimized constellations obtained from SEP-A, MI-A, MI-B, and APSK-A, and compare their performances with that of conventional constellations (QAM and PSK) and those proposed in prior work. We remark that many of the constellations proposed in prior work for phase noise have a well-defined structure, and they can be optimized using the analytical framework provided in SEP-A, MI-A, and MI-B for a given $E_{\tr b}/N_{0}$ and $\sigma_{\tr{p}}^{2}$. However, for our comparative study, we do not optimize these structures for a given $E_{\tr b}/N_{0}$ and $\sigma_{\tr{p}}^{2}$.

\begin{figure*}[!t]
\begin{center}
\begin{tabular}{cc}
\includegraphics[width = 3.5in, keepaspectratio=true]{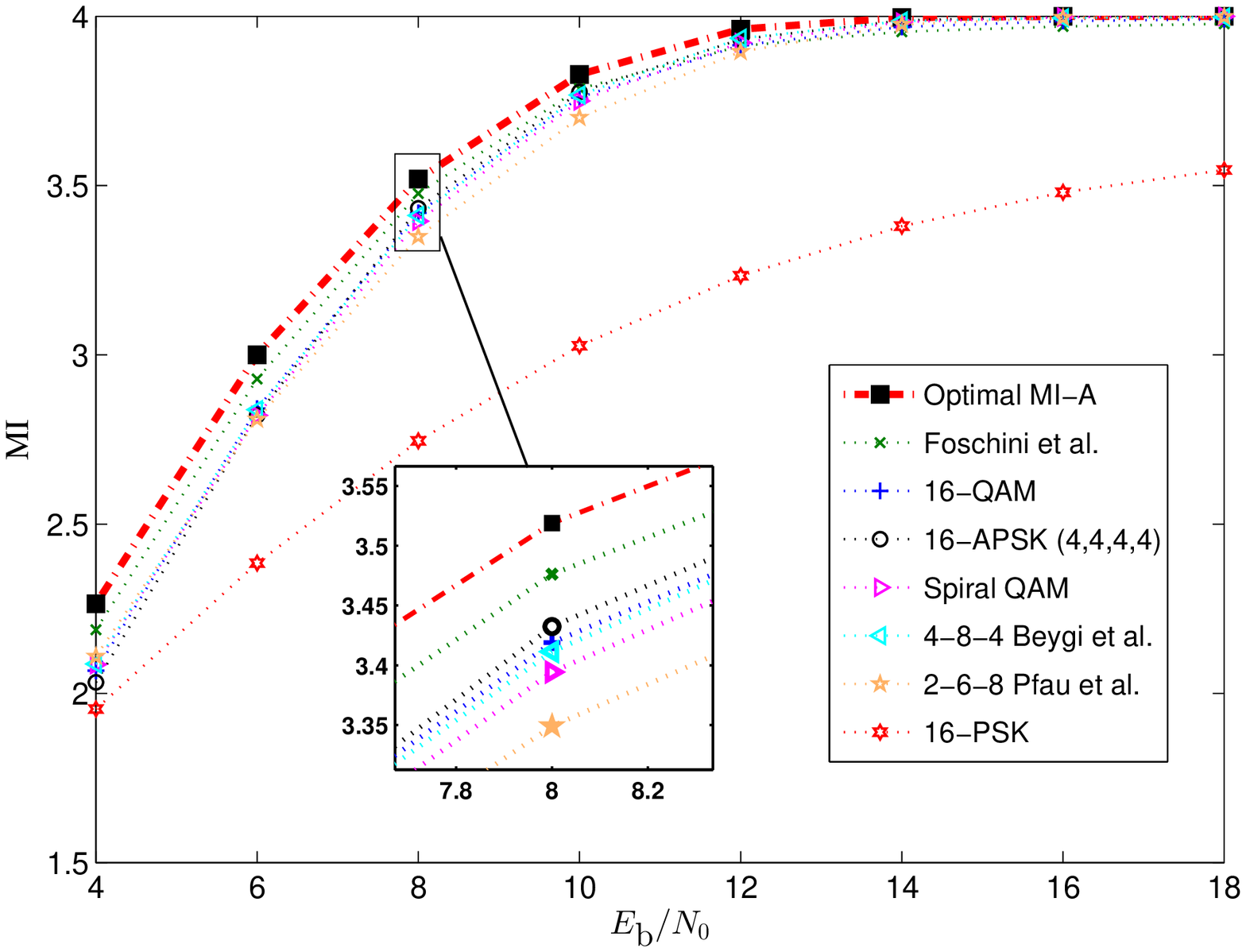}
&
\includegraphics[height = 2.85in, width = 3.5in]{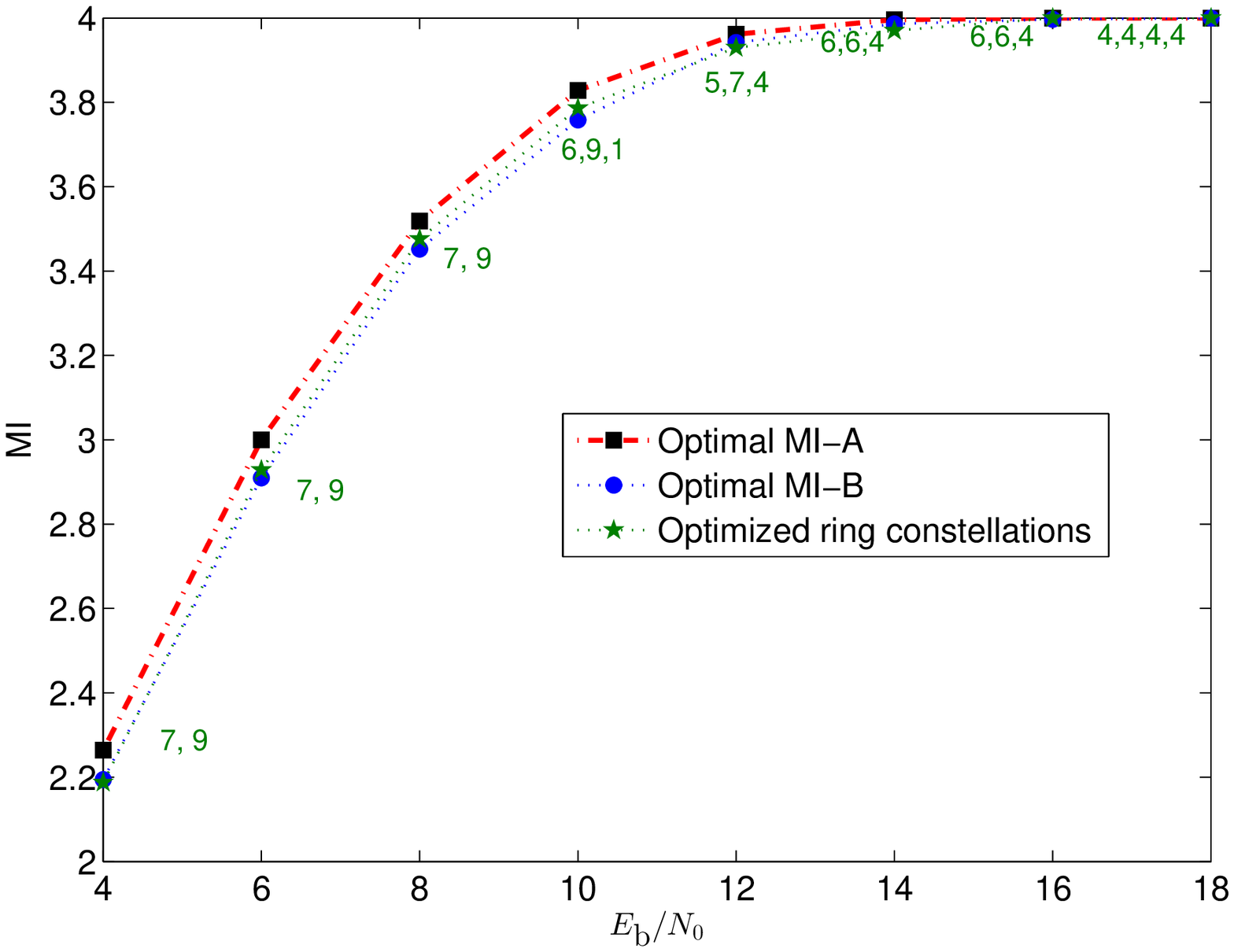}
\\
(a) & (b)\\
\end{tabular}
\caption{(a) MI with GAP-D \refeq{eq:mi_mldet1} for different constellations for $\sigma_\tr{p}^{2} = 0.01\, \text{rad}^{2}$.(b) MI with GAP-D \refeq{eq:mi_mldet1} for MI-A optimized, MI-B optimized, and the optimized ring constellations (along with the optimal ring configurations) obtained by optimizing $I_{\text{DD}}(A^{(i)})$.}\label{fig:mia_comp}
\end{center}
\end{figure*}

\begin{figure*}[!ht]
\begin{center}
\begin{tabular}{cc}
\includegraphics[width = 3.5in, keepaspectratio=true]{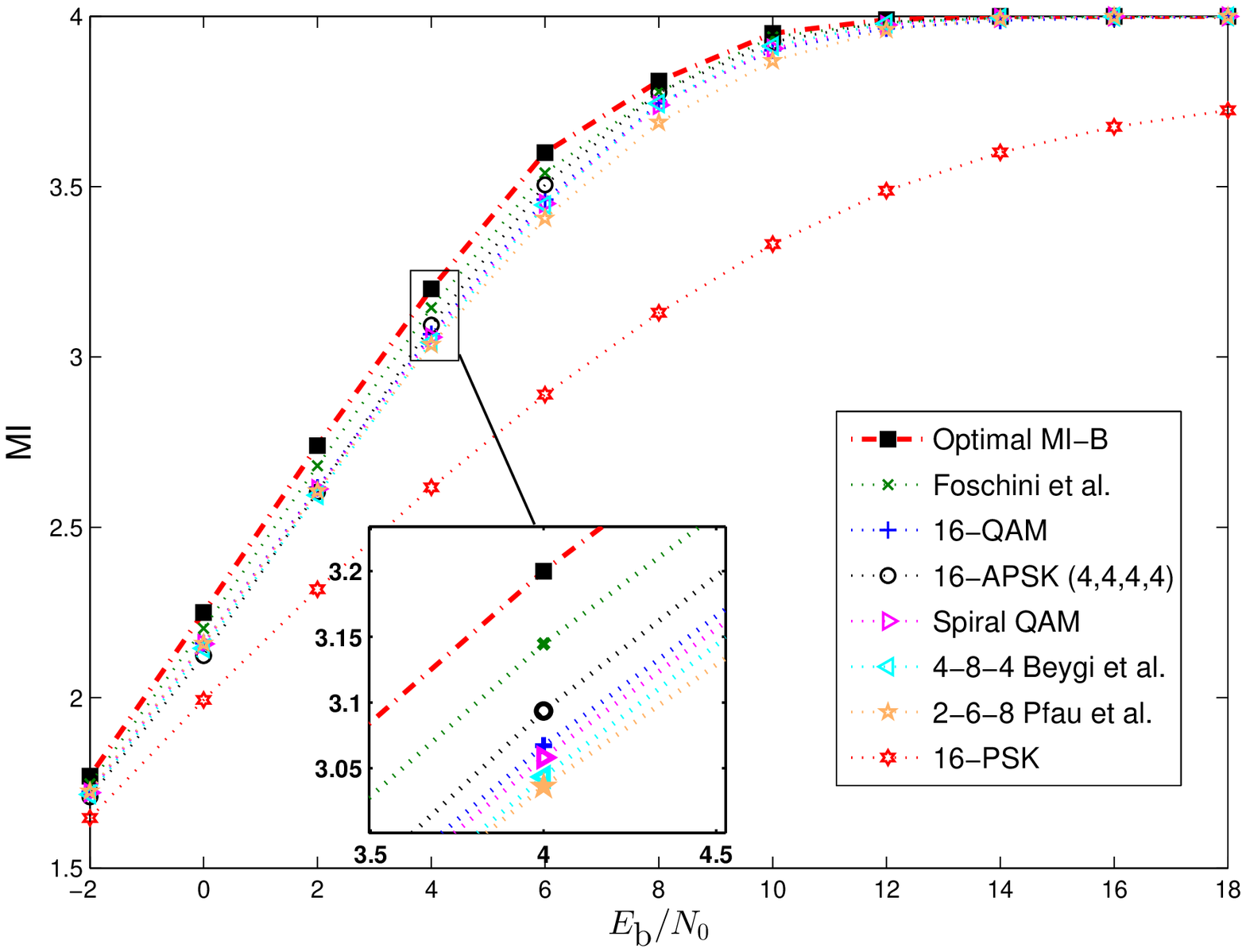}
&
\includegraphics[width = 3.5in, keepaspectratio=true]{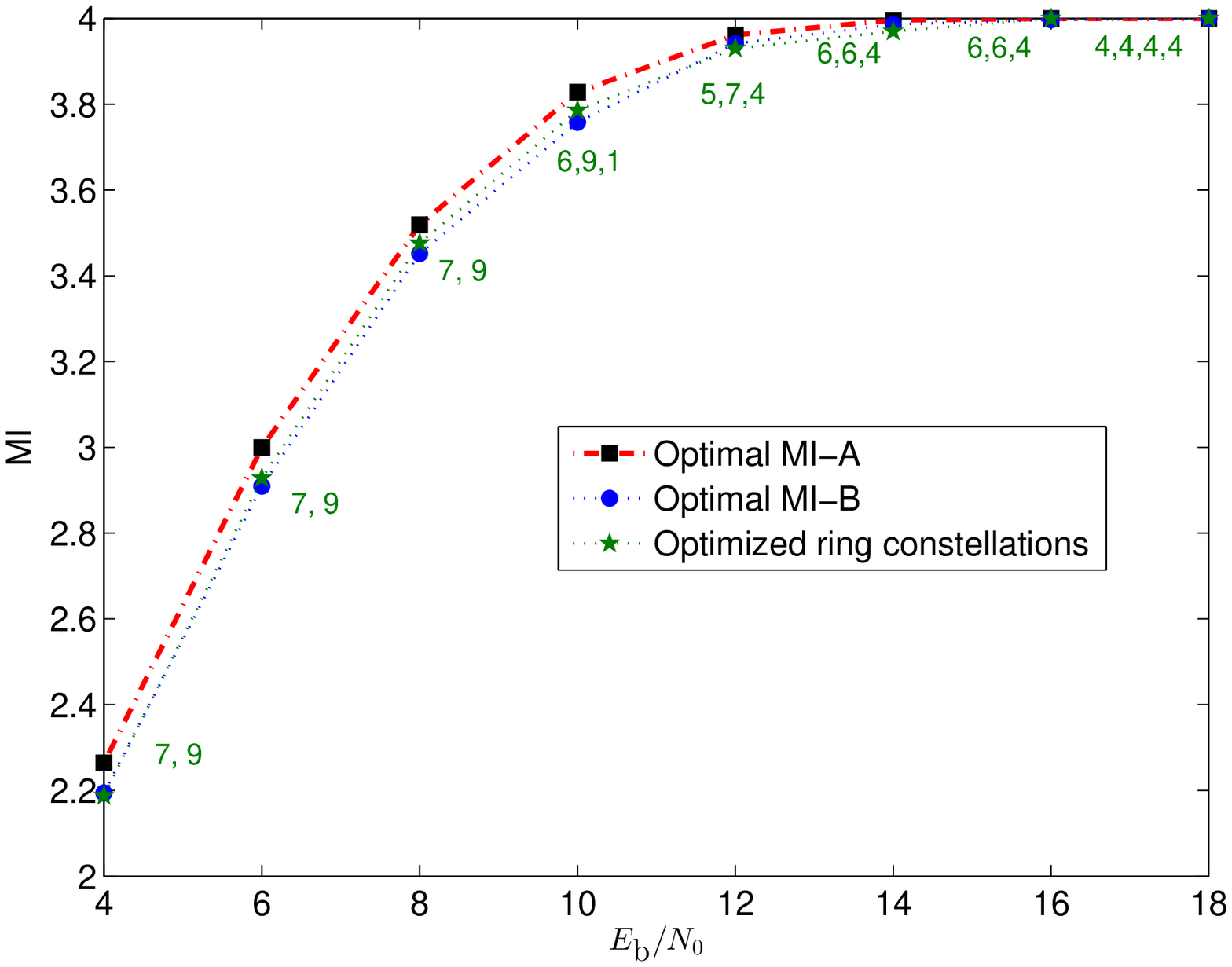}
\\
(a) & (b)\\
\end{tabular}
\caption{(a) The actual MI  \refeq{eq:mi_mlmemoryless} for different constellations for $\sigma_\tr{p}^{2} = 0.01\, \text{rad}^{2}$. (b) The actual MI for MI-B optimal, MI-A optimal and the optimized ring constellations (along with the optimal ring configurations) obtained by optimizing $I_{\text{DC}}(A^{(i)})$.}\label{fig:mib_comp}
\end{center}
\end{figure*}

\subsection{Comparison of SEP and MI with GAP-D}

We first evaluate the SEP with GAP-D of the optimized constellations in Fig. \ref{fig:sep_comp}(a) for different values of $E_{\tr b}/N_{0}$ ($4$--$20$ dB), and a fixed phase noise variance  $\sigma_\tr{p}^{2} = 0.01\, \text{rad}^{2}$. At low $E_{\tr b}/N_{0}$ ($4$--$8$ dB), we observe that the  $(1,6,9)$-APSK configuration proposed by Foschini \emph{et al.} \cite{fos73} performs slightly better than all other constellations considered. The performance of the optimal constellations obtained from SEP-A is slightly worse relative to the $(1,6,9)$-APSK configuration in this scenario. This is because the SEP derived in \refeq{eq:p_error} is inaccurate for low values of $E_{\tr b}/N_{0}$, and hence the optimization formulation is suboptimal in such scenarios.

As $E_{\tr b}/N_{0}$ increases, the optimized constellations obtained from SEP-A outperform all other constellations. For medium-to-high $E_{\tr b}/N_{0}$ ($10$--$20$ dB), amongst constellations known from prior work, the spiral constellation \cite{kwa08} and $(4,4,4,4)$-APSK \cite{li08} perform the best. Their performance is observed to be only around $1$ dB from that of the optimized constellations. This can be attributed to the large number of energy levels in a spiral constellation and $(4,4,4,4)$-APSK constellation. Also, it can be seen that equal-energy points in $(4,4,4,4)$-APSK  are separated by large angular distances. These factors also play a decisive role in determining the asymptotic SEP (or error floors) of constellations. For the given phase noise scenario and almost all $E_{\tr b}/N_{0}$ values considered, PSK and QAM (in the same order) are observed to be the worst performing constellations. In Fig. \ref{fig:sep_comp}(b), we plot the performance of the SEP-A optimal constellations against the APSK constellations obtained by using SEP as the criterion in the formulation APSK-A. We observe that the optimal APSK-A constellations have comparable SEP performance with respect to that of those optimizing SEP-A for the values of $E_{\tr b}/N_{0}$ and $\sigma_\tr{p}^{2}$ considered. {To evaluate the robustness of the optimized constellations to varying $E_b/N_0$, we also plot in the figure the performance of the SEP-A constellation optimized for $E_{\tr b}/N_{0} = 20$ dB. Below 18 dB, we observe a loss with respect to the constellations optimized for each $E_b/N_0$. Thus, we can conclude that the optimized constellations are quite sensitive to varying $E_{\tr b}/N_{0}$.}

In Fig. \ref{fig:mia_comp}(a), the MI of different constellations for the effective channel with memoryless phase noise and GAP-D is presented. We observe that the MI of the optimal constellations from MI-A outperform all other constellations from prior work for all $E_{\tr b}/N_{0}$ values. At low $E_{\tr b}/N_{0}$ ($4$--$8$ dB), the performance of the optimized constellations is closely followed by that of $(1,6,9)$-APSK. As the $E_{\tr b}/N_{0}$ is increased, we observe that the MI of the optimized constellations and those proposed in prior work become comparable. For all $E_{\tr b}/N_{0}$ values, the performance of PSK is seen to be the worst, which is expected given its poor SEP. Also, for low values of $E_{\tr b}/N_{0}$, the performance of $(4,4,4,4)$-APSK constellation is slightly worse than the other constellations proposed for phase noise scenarios in prior work. In Fig. \ref{fig:mia_comp}(b), we compare the performance of the optimal constellations from MI-A with that of the optimal APSK constellations obtained from using $I_{\text{DD}}(A^{(i)})$ in formulation APSK-A. Once again, we observe that the optimal APSK constellations for this criterion are also the optimizers for the SEP criterion.

\subsubsection{Asymptotic SEP Performance}
In Fig. \ref{fig:sep_comp}(a), we observe that conventional constellations and those proposed in prior work suffer from an error floor in their SEP performance at high $E_{\tr b}/N_{0}$. This behavior is caused by symbol points that are of equal energy, and are not separated by large angular distances \refeq{eq:error_floor}. Thus, it can be inferred that a constellation with symbol points all having different energy levels will not have any error floor. Furthermore, if some of the points in the constellation have the same energy, then they should be separated by a large angular distance in order to reduce the error floor. In Table \ref{table:err_flr}, we present the error floor of different constellations from prior work for a phase noise variance $\sigma_\tr{p}^{2} = 0.01 \text{ rad}^{2}$. It can be seen that the floor appears at undesirably high values, particularly for QAM and PSK modulations, while constellations optimized for high $E_{\tr b}/N_{0}$ do not suffer from it. One way to reduce the floor level of these constellations is to translate their I-Q axes. For example, traditionally $16$-PSK and $16$-QAM have their origin at $(0,0)$, which can be translated to $(0.5,0.3)$ (before normalizing the constellation to the power constraint $P$) to have no error floor for $\sigma_\tr{p}^{2} = 0.01$. The process of translating the origin of these constellations helps converting the equal energy points to non-equal energy points.

%Note that when the $E_{\tr b}/N_{0}$ tends to infinity (i.e., where there is non-zero AWGN in the channel), we conjecture that the optimal constellation set is a set of points where no two symbol points have the same energy and the energy levels are maximally separate.

\vspace{16mm}
\begin{table}[t]
\caption{Error Floors for Different Constellations $\sigma_\tr{p}^{2} = 0.01 \text{ rad}^{2}$ } % title of Table
\centering % used for centering table
\begin{tabular}{l c} % centered columns (4 columns)
\hline %inserts double horizontal lines
Constellation &  SEP  \\ %[0.5ex] % inserts table
%heading
\hline % inserts single horizontal line
Optimal Constellation at $40$ dB  & $0$ \\
$16$-PSK & $0.0498$   \\
$16$-QAM & $3.5\times 10^{-4}$   \\
Foschini \emph{et al.} $(1,6,9)$ & $2.2\times 10^{-4}$   \\
$16$-Spiral QAM \cite{kwa08} & $ 4.7 \times 10^{-15}$   \\
$16$-APSK $(4,4,4,4)$  & $4.7 \times 10^{-15}$  \\
%$16$-Elliptical [1] $(4,4,4,4)$  & $4.5 \times 10^{-15}$ & $1.9\times10^{-28}$  \\
Beygi \emph{et al.} (Optimized $4,8,4$)  & $5.3\times 10^{-5}$   \\
Pfau \emph{et al.} (Optimized $2,6,8$)  & $5.2 \times 10^{-5}$   \\
%[.5ex] % [1ex] adds vertical space
\hline %inserts single line
\end{tabular}
\label{table:err_flr} % is used to refer this table in the text
\end{table}

\subsection{Comparison in terms of MI of Memoryless Phase Noise Channel}

We compare the MI of the optimized constellations obtained from MI-B with those from prior work and other conventional constellations for different values of $E_{\tr b}/N_{0}$ (-$2$--$20$ dB), and a fixed phase noise variance  $\sigma_\tr{p}^{2} = 0.01\, \text{rad}^{2}$ in Figs. \ref{fig:mib_comp}(a) and (b). For a given phase noise variance, the gain in MI rendered by the optimized constellations for the memoryless phase noise channel is significant for low $E_{\tr b}/N_{0}$ values. As $E_{\tr b}/N_{0}$  is increased, we observe that this gap decreases, and almost all constellations have comparable MI with respect to the optimized constellations. Amongst the known constellations, it can be seen that $(4,4,4,4)$-APSK performs slightly worse than the other constellations at low $E_{\tr b}/N_{0}$. However as $E_{\tr b}/N_{0}$  is increased, its MI becomes comparable with that of the other constellations. A similar observation was made for this constellation in the MI-A formulation. In the case of $16$-PSK, we observe that its MI is much worse than that of all the other constellations considered.

In Fig. \ref{fig:mib_comp}(b), we compare the performance of the  MI-B  optimal constellations with the optimized ring constellations obtained by using $I_{\text{DC}}(A^{(i)})$ as the objective function in formulation APSK-A. We observe that the performance of these sets of constellations is close to each other. %We observe that the MI of the MI-A optimal constellations for the channel with GAP-D \refeq{eq:mi_mldet1} is significantly lower than that of the MI-B optimal constellations for the memoryless phase noise channel \refeq{eq:mi_mlmemoryless}.
Further, we observe that the optimized constellations obtained from SEP-A have MI close to that of the optimized constellations obtained from MI-B. Likewise, in Figs. \ref{fig:sep_comp}(b) and \ref{fig:mia_comp}(b), we observe that constellations that maximize MI for the memoryless phase noise channel achieve SEP  performance close to that of the optimized constellations obtained from SEP-A and MI-A.

\section{Conclusions}
\label{sec:conc}

We presented an analytical framework for constellation design for phase noise channels based on three optimization formulations. We first designed constellations that minimize the symbol error probability of the maximum likelihood detector. The optimized constellations perform better than all known constellations for moderate-to-high SNRs. Furthermore, they do not suffer from an error floor. Next, we designed constellations that maximize the mutual information of the effective (discrete) channel consisting of memoryless phase noise, AWGN and the ML detector. The optimized constellations outperform all other constellations from prior work for all SNRs and phase noise variances. Finally, we optimized constellations to maximize the MI of a memoryless phase noise channel. We provided two analytical characterizations for the MI, based on a low instantaneous phase noise approximation, and on a high SNR approximation, respectively. Compared to state-of-the-art constellations, the gain in MI yielded by the optimized constellations is more pronounced at low SNR. %\textcolor{red}{In our work, we did not consider the problem of jointly optimizing the constellations and their labeling in order to maximize the GMI, which is relevant to coded systems that use BICM schemes, and we remark that this is an interesting direction for future work. }

We also constructed simple ring (APSK) constellations, which achieve comparable performances with respect to that of the constellations obtained from the global optimization formulations.

\bibliographystyle{IEEEbib}

\end{document}